\documentclass[prd,amsmath,amssymb,groupedaddress,letterpaper,twocolumn,nofootinbib]{revtex4}
\usepackage{graphicx} 

\renewcommand{\a}{\alpha}
\renewcommand{\b}{\beta}

\newcommand{\cD}{{\cal D}}
\newcommand{\cH}{{\cal H}}

\newcommand{\cL}{{\cal L}}
\newcommand{\cM}{{\cal M}}

\renewcommand{\d}{\delta}

\newcommand{\nn}{\nonumber}
\newcommand{\p}{\partial}

\newcommand{\D}{\not\!\!D}

\renewcommand{\Re}{\text{Re\,}}
\renewcommand{\Im}{\text{Im\,}}
\newcommand{\tr}{\text{tr\,}}
 
\renewcommand{\Phi}{\varPhi}
\renewcommand{\Psi}{\varPsi}
\renewcommand{\Omega}{\varOmega}
\renewcommand{\Gamma}{\varGamma}

\begin{document}

\title{The null energy condition and instability}
 
\author{Roman~V.~Buniy} \email{roman@uoregon.edu}
\author{Stephen~D.~H.~Hsu} \email{hsu@duende.uoregon.edu}
\author{Brian~M.~Murray} \email{bmurray1@uoregon.edu}

\affiliation{Institute of Theoretical Science, University of Oregon,
  Eugene OR 94703-5203}

\begin{abstract}
We extend previous work showing that violation of the null energy
condition implies instability in a broad class of models, including
gauge theories with scalar and fermionic matter as well as any perfect
fluid. Simple examples are given to illustrate these results. The role
of causality in our results is discussed.  Finally, we extend the
fluid results to more general systems in thermal equilibrium. When
applied to the dark energy, our results imply that $w = p / \rho$ is
unlikely to be less than $-1$.
\end{abstract}
 
\maketitle

\section{Introduction} \label{introduction}

Every spacetime is a possible solution to Einstein's equations
for some particular choice of energy-momentum tensor $T_{\mu
\nu}$. It is therefore important to understand what general
restrictions can be placed on the energy-momentum of a physical
system. Such restrictions are referred to as energy conditions, and
play an important role in general relativity. In an earlier paper
\cite{buniy}, two of the present authors demonstrated a direct
connection between stability and the null energy condition (NEC)
\cite{null}, $T_{\mu \nu} n^\mu n^\nu \ge 0$ for any null vector
$n^\mu$ (satisfying $g_{\mu\nu}n^\mu n^\nu = 0$). The main results of
the study in Ref.~\cite{buniy} are: (1) classical solutions of both
minimally and non-minimally coupled scalar-gauge models which violate
the NEC are unstable, (2) a quantum state in which the expectation of
the energy-momentum tensor violates the NEC cannot be the ground state
(includes models with fermions), (3) perfect fluids which violate the
NEC are unstable. These results suggest that violations of the NEC in
physically interesting cases are likely to be ephemeral.

Using the Raychaudhuri equation for the expansion of a hypersurface
null congruence, one can show that null geodesics have non-increasing
expansion as long as the NEC is obeyed \cite{Hawking-Ellis}. This
result plays an important role in the classical singularity theorems
of general relativity \cite{Hawking-Ellis} and in proposed covariant
entropy bounds \cite{bousso}. It also implies that NEC violating
matter is necessary for the construction of Lorentzian
wormholes~\cite{wormholes}.
 
This work is an extension of Ref.~\cite{buniy}. We give some explicit
examples of the classical stability analysis (Sec. \ref{classical}),
additional details regarding the local instability in quantum states
which violate the NEC (Sec. \ref{quantum}), a simplified and more
general proof of the result that the fermionic effective action does
not lead to violation of the NEC (Sec. \ref{fermions}), and a
generalization of our analysis of fluids to any system in thermal
equilibrium (Sec. \ref{fluids}). In Ref.~ \cite{dubovsky} it was shown that
models with superluminal modes can evade our results in some
cases. However, in Ref.~\cite{nima} it was noted that such models are
acausal. In the appendix, we show that in causal models, the
NEC is sufficient to determine stability. In Sec. \ref{classical} it is
shown that considerations involving causality or superluminality are
necessary only in cases where the classical background violates all
rotational symmetries. Cosmological models of the dark energy must
exhibit isotropy and hence are covered by the earlier results of
Ref.~\cite{buniy}.

The NEC has direct implications for the dark energy equation of state,
often given in terms of $w=p/\rho$. Dark energy has positive energy
density $\rho$ and the energy-momentum tensor is $T_{\mu \nu} = \text
{diag\,}(\rho, p, p, p)$ in the comoving cosmological
frame. Therefore, $w < -1$ implies violation of the NEC. Instability
as a consequence of $w < -1$ was studied previously in scalar
models~\cite{instability} and in general in Ref.~\cite{buniy}. The
analysis applies to dark energy models such as
k-essence~\cite{k-essence}, rolling tachyon condensate~\cite{tachyon},
and phantoms~\cite{phantom}.

\section{Classical field theories}\label{classical}

In this section we will first look at some examples and then derive
the general result of Ref.~\cite{buniy}, which relates the violation
of the NEC to classical instabilities.  Our metric convention is such
that in a locally inertial frame $g_{\mu\nu}
=\text{diag\,}(1,-1,\ldots,-1)$.

\subsection{Scalar field model} \label{scalar}

Consider the action
\begin{eqnarray}
  S =\int d^dx\,|g|^\frac{1}{2}\left[Q(X)-V(\Phi)\right],
    \label{phi:S}
\end{eqnarray}
where $X=\phi_{a,\mu}\phi^{a,\mu}$ and $\Phi=\phi_a\phi^a$.
Throughout, commas indicate partial derivatives.  By setting the
variation of the action to zero, we find the equations of motion for
the fields $\phi_a$:
\begin{eqnarray}
  \nabla_\mu(Q_X \phi^{a,\mu}) = -V_{\Phi}\phi^{a}. \label{phi:eom}
\end{eqnarray}
In what follows, we assume that we have found
a solution $\phi_a(x)$ to these equations of motion, and
the quantities $Q_X$, etc. are evaluated at this solution.

\subsubsection{Stability}

In order to determine whether the solution $\phi_a(x)$ to the
equations of motion is stable against small fluctuations $\d\phi_a$,
we consider the second variation of the action,
\begin{eqnarray}
  \d^2S = \int d^dx\,|g|^\frac{1}{2}\left(\delta^2 Q -\d^2 V
    \right),\label{phi:d2S}
\end{eqnarray}
where
\begin{eqnarray}
    \delta^2 Q &=&\left(2Q_X\delta^{ab}g^{\mu\nu}
    +4Q_{XX}\phi^{a,\mu}\phi^{b,\nu}\right)
    \d\phi_{a,\mu}\d\phi_{b,\nu},\label{phi:d2F}\\ \delta^2 V
    &=&\left(2V_\Phi\delta^{ab}+4V_{\Phi\Phi}\phi^a\phi^b\right)
    \delta\phi_a\delta\phi_b.\label{phi:d2V}
\end{eqnarray}

We now use a locally inertial frame. For $\delta^2\cL$, the canonical
momentum is
\begin{eqnarray}
  \d\pi^a =
    4Q_X\d\phi^{a,0}+8Q_{XX}\phi^{a,0}\phi^{b,\nu}\d\phi_{b,\nu}
    \label{phi:dpi}
\end{eqnarray}
and the Hamiltonian is $\d^2\cH =\d^2 K +\d^2 V$. Here the kinetic
term is
\begin{eqnarray}
  \d^2 K &=& (2Q_X\d^{ab}+4Q_{XX}\phi^{a,0}\phi^{b,0})
    \d\phi_{a,0}\d\phi_{b,0} \nn\\ &+&
    (2Q_X\d^{ab}\d^{ij}-4Q_{XX}\phi^{a,i}\phi^{b,j})\
    d\phi_{a,i}\d\phi_{b,j},
\end{eqnarray}
where $\delta\phi_{a,0}$ are found by solving Eq.~(\ref{phi:dpi}),
\begin{eqnarray}
  \delta\phi_{a,0}&=&
    \frac{1}{4Q_X}\left({\delta^b}_a-\frac{4Q_{XX}\phi^{b,0}\phi_{a,0}}
    {2Q_X+4Q_{XX}\phi_{c,0}\phi^{c,0}}\right) \nn \\ &\times&
    \left(\delta\pi_b-8Q_{XX}
    \phi_{b,0}\phi^{c,i}\delta\phi_{c,i}\right).
\end{eqnarray} 

\begin{figure}
\includegraphics[width=6cm]{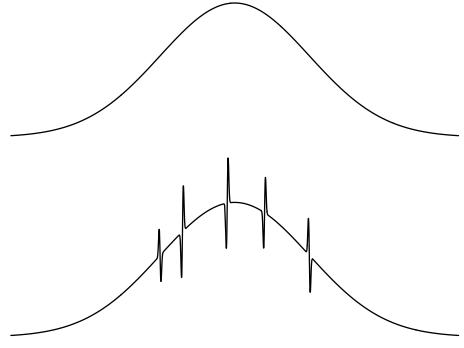}
\caption{In a model that violates the NEC, an initially smooth 
field configuration (top) is unstable. The system can lower its
total energy by developing gradients (bottom), thereby decreasing its
kinetic energy, while leaving its potential energy unchanged.}
\label{figure_grad}
\end{figure}

We are interested in finding a sufficient condition for the solutions
of this model to be unstable.
Moreover, we are free to investigate any set of variations we like.
In particular, let us consider a set of
variations such that $\d{\phi}_{1,\mu} \neq 0$ and
$\d{\phi}_{a,\mu}=0$ ($a\neq 1$). In order to obtain a simple condition
for instability, we further restrict $d-1$ quantities
$\d{\phi}_{1,i}$ to satisfy the following equation:
\begin{eqnarray}
  ({\phi}^{1,0}\d{\phi}_{1,0})^2 =({\phi}^{1,i}\d{\phi}_{1,i})^2.
    \label{phi:cond}
\end{eqnarray}
For this choice of variations,
\begin{eqnarray}
  \d^2 K=Q_X\left[(\delta{\phi}_{1,0})^2
    +\sum_{i}(\delta{\phi}_{1,i})^2\right]. \label{phi:d2K}
\end{eqnarray}
We see that the classical solution $\phi_a(x)$ is unstable for $Q_X
<0$. Specifically, it is unstable to the formation of gradients
because the system can lower its kinetic energy by creating these
gradients, while leaving the potential energy fixed. See
Fig.~\ref{figure_grad} for a sketch.

\subsubsection{NEC}

Now let us consider the NEC for the model~(\ref{phi:S}). The NEC
states that $T_{\mu\nu}n^\mu n^\nu \geq 0$ for any null vector
$n^\mu$. The energy-momentum tensor that couples to gravity is
\begin{eqnarray}
  T_{\mu\nu} =-\cL g_{\mu\nu} +2\cL_{g^{\mu\nu}}.
\end{eqnarray}
Computing the derivative $\cL_{g^{\mu\nu}} =Q_X
{\phi^a}_{,\mu}\phi_{a,\nu}$, we see that the NEC requires
\begin{eqnarray}
  Q_X \sum_{a}(n^\mu\phi_{a,\mu})^2 \geq 0,\label{phi:near_nec}
\end{eqnarray}
and therefore,
\begin{eqnarray}
  Q_X \geq 0. \label{phi:nec}
\end{eqnarray}
Comparing Eqs.~(\ref{phi:d2K}) and (\ref{phi:nec}), we conclude that
the violation of the NEC in this model implies that the classical
solutions to the equations of motion are unstable to the formation of
gradients.

\subsection{Gauge field model} \label{gauge}

Consider the action
\begin{eqnarray}
  S =\int d^4x\,|g|^\frac{1}{2}\left[Q(Y)+\theta|g|^{-\frac{1}{2}}
     \epsilon^{\mu\nu\rho\sigma} F_{\mu\nu}F_{\rho\sigma} \right],
    \label{A:S}
\end{eqnarray}
where $\theta$ is a dimensionless parameter, $|g|^{-\frac{1}{2}}
\epsilon^{\mu\nu\rho\sigma}$ is the totally antisymmetric tensor in
curved spacetime, $Y = F_{\mu\nu}F^{\mu\nu}$, and
$F_{\mu\nu}=A_{\nu;\mu}-A_{\mu;\nu}=A_{\nu,\mu}-A_{\mu,\nu}$.  A
semicolon indicates the covariant derivative appropriate for curved
space, and the last equality holds because the connection has no
torsion.  Classical electrodynamics corresponds to
$Q(Y)=-\tfrac{1}{4}Y$ and $\theta=0$.

As before, the equations of motion are found by setting the variation
of the action to zero and integrating by parts. Here the equations of
motion for $A_\mu$ are
\begin{eqnarray}
  \nabla_\mu(Q_YF^{\mu\nu}+\theta |g|^{-\frac{1}{2}}
    \epsilon^{\mu\nu\rho\sigma}F_{\rho\sigma}) =0. \label{A:eom}
\end{eqnarray}
In what follows we will assume that we have found a solution
$A_\mu(x)$ to these equations of motion.

\subsubsection{Stability}

As in the stability analysis for the scalar field model of the
previous section, we consider the second variation of the action.
Since the Lagrangian of (\ref{A:S}) has no dependence on the gauge field
$A_\mu$, this is simply
\begin{eqnarray}
  \d^2 S=\int d^dx\,|g|^\frac{1}{2}\cL_{A_{\a;\mu}A_{\b;\nu}} \delta
    A_{\a;\mu}\delta A_{\b;\nu}. \label{A:d2S}
\end{eqnarray}
Derivatives of $\cL$ with respect to the covariant derivative are
equal to those with respect to the regular derivative, which in turn
are proportional to derivatives with respect to the field strength,
$\cL_{A_{\a;\mu}A_{\b;\nu}}=\cL_{A_{\a,\mu} A_{\b,\nu}} =4\cL_{F_{\a
\mu}F_{\b \nu}}$.  Antisymmetry of $F_{\a \mu}$ is used to show the
second equality.  Therefore, to evaluate Eq.~(\ref{A:d2S}), we compute
the derivatives of the Lagrangian with respect to the field strength.
Again using the antisymmetry of $F_{\a \mu}$:
\begin{eqnarray}
  \cL_{F_{\a \mu}F_{\b \nu}} = M^{\a \b}g^{\mu \nu}+N^{\a \mu \b \nu},
\end{eqnarray}
where we have defined
\begin{eqnarray}
  M^{\a\b} &=& 4Q_Yg^{\a\b}, \\ N^{\a\mu\b\nu} &=&
  -4Q_Yg^{\a\nu}g^{\b\mu} +16Q_{YY}F^{\a\mu}F^{\b\nu} \nn\\
  && +8\theta|g|^{-\frac{1}{2}} \epsilon^{\a\mu\b\nu}. \label{A:N}
\end{eqnarray}

The second variation of the Lagrangian now becomes
\begin{eqnarray}
  \delta^2\cL &=&\left(M^{\a\b}g^{\mu\nu} +N^{\a\mu\b\nu}
    \right)\delta A_{\a;\mu}\delta A_{\b;\nu}.\label{A:d2L}
\end{eqnarray}
For the Lagrangian~(\ref{A:d2L}) the canonical momentum is
\begin{eqnarray}
  \d\pi^\a =2M^{\a\b}\d A_{\b;0}+2N^{\a0\b\nu}\d A_{\b;\nu},
    \label{A:pi}
\end{eqnarray}
where we use a locally inertial frame.  This allows us to write
\begin{eqnarray}
  \d^2\cH &=& \left(M^{\a\b}+N^{\a0\b0}\right) \d A_{\a;0}\d A_{\b;0}
    \nn\\ &+&\left( M^{\a\b}\d^{ij} -N^{\a i\b j}\right) \d A_{a;i}\d
    A_{\b;j},
\end{eqnarray}
and $\d A_{\a;0}$ are related to $\d \pi_{\a}$ and $\d A_{\a;i}$
through Eq.~(\ref{A:pi}).  To show that this model can be unstable, we
choose variations such that $\delta{A}_{1;\mu}\not =0$ and
$\delta{A}_{\a;\mu}=0$ $(\a \neq 1)$.  We further restrict $d-1$
quantities $\delta{A}_{1;i}$ to satisfy the following equation:
\begin{eqnarray}
  {N}^{1010}\delta{A}_{1;0} \delta{A}_{1;0} ={N}^{1 i1
    j}\delta{A}_{1;i}\delta{A}_{1;j}.
    \label{A:cond}
\end{eqnarray}
Therefore,
\begin{eqnarray}
  \d^2\cH&=&M^{11} \d A_{1;0}\d A_{1;0}+ M^{11}\d^{ij}\d A_{1;i}\d
    A_{1;j} \nn\\ &=&-4Q_Y\left[(\delta{A}_{1;0})^2
    +\sum_{i}(\delta{A}_{1;i})^2\right]. \label{A:d2K}
\end{eqnarray}
We see that if $Q_Y>0$ there is the same type of instability as in the
scalar field case, independent of the value of $\theta$ in the
$F\tilde{F}$ term.  Note that in this model the total energy is simply
the kinetic energy because there is no potential energy, i.e., no
dependence on the field $A_\mu$.

\subsubsection{NEC}

In order to determine whether this model violates the NEC, we compute
the following derivative
\begin{eqnarray}
  \cL_{g^{\mu\nu}} = \tfrac{1}{2}M^{\a\b}F_{\a\mu}F_{\b\nu}
      +\tfrac{1}{2}\theta g_{\mu\nu} \vert g\vert^{-\frac{1}{2}}
      \epsilon^{\a\b\rho\sigma}F_{\a\b}F_{\rho\sigma},
      \label{gauge:lg}
\end{eqnarray}
where we have used $\p \vert g\vert/\p g^{\mu\nu}=-|g|g_{\mu\nu}$.
From Eq.~(\ref{gauge:lg}), we see that for this model the NEC requires
\begin{eqnarray}
  Q_Y \Psi_{\a}\Psi^{\a} \geq 0,\label{gauge:nec}
\end{eqnarray}
where $\Psi_\a = n^\mu F_{\a\mu}$.  Note that the $F\tilde{F}$ term
does not enter into the NEC.  We can determine the sign of
$\Psi_{\a}\Psi^{\a}$ as follows.  First notice
\begin{eqnarray}
  n^\a \Psi_\a = n^\a n^\mu F_{\a\mu} = 0, \label{A:npsizero}
\end{eqnarray}
by antisymmetry of $F_{\a\mu}$.  Now use a locally inertial frame, and
let $\Psi_{\a}=(\Psi_0,\mathbf{\Psi})$ and
$n_{\a}=(|\mathbf{n}|,\mathbf{n})$. Then Eq.~(\ref{A:npsizero})
implies
\begin{eqnarray}
  n^\a \Psi_\a = |\mathbf{n}|\Psi_0 -|\mathbf{n}| |\mathbf{\Psi}|
    \cos \varphi = 0,
\end{eqnarray}
which means that $\Psi_{\a}$ is non-timelike, $\Psi_{\a}\Psi^{\a} \leq
0$. Therefore, Eq.~(\ref{gauge:nec}) implies that the NEC is violated
if $Q_Y > 0$. As noted below Eq.~(\ref{A:d2K}), $Q_Y > 0$ is also a
sufficient condition for the model to be unstable. Thus, the violation
of the NEC in this model implies that the classical solutions to the
equations of motion are unstable to the formation of gradients.

\subsection{General case}

We now give an extended version of the general proof of the classical
field theory result found in Ref.~\cite{buniy}.  Consider a theory of
scalar, $\phi_a$, and gauge, $A_{a\alpha}$, fields in a fixed
$d$-dimensional spacetime with the metric $g_{\mu\nu}$.~\footnote{In
our analysis we assume a fixed, but possibly non-trivial, spacetime
metric $g_{\mu\nu}$. We study local instabilities of the matter fields
propagating on that spacetime background that would manifest
themselves on microphysical timescales. It is always possible to work
in an inertial frame at any particular point in spacetime. One might
wonder whether backreaction of the metric might dampen the growth of
an instability. Since the principle of equivalence (which guarantees
the existence of an inertial frame) applies everywhere, it is easy to
see that stability cannot be restored unless the null energy condition
ceases to be violated. That is, although backreaction could in
principle decrease the magnitude of the negative eigenvalue governing
the instability, the eigenvalue cannot go to zero or change sign
unless the null energy condition is no longer violated. Furthermore,
the amount by which backreaction might decrease the growth of the
instability is suppressed by the Planck scale (this is easy to deduce
from the Einstein equations), so it is negligible except in circumstances
in which quantum gravitational effects are important, and known
methods cease to apply.} We limit ourselves to theories whose
equations of motion are second order differential equations, so the
Lagrangian density $\cL$ for the system is assumed to depend only on
the the scalar fields $\phi_a$, their covariant derivatives
$D_\mu\phi_a$, and the gauge field strengths $F_{a\mu\nu}$. The
scalars may transform in any representation of the gauge group. We
impose Lorentz invariance on $\cL$, but do not require overall gauge
invariance. That is, we allow for fixed tensors with gauge indices
(but no Lorentz indices) which can be contracted with the fields, as
is the case, e.g., in models with spontaneous symmetry breaking.

Although we consider minimal coupling between matter fields and
gravity, non-minimal models of the type below are related to minimal
models by conformal transformation~\cite{instability}. Indeed,
consider the non-minimally coupled action
\begin{eqnarray}
  S=\int
  d^dx\,|g|^\frac{1}{2}[\tfrac{1}{2}B(\phi_a)R
    +\cL(\phi_a,D_\mu\phi_a,F_{a\mu\nu})],\label{FT:Snm}
\end{eqnarray}
where $R$ is the scalar curvature. For a single minimally and
non-minimally coupled scalar, $B=1$ and $B=1- \tfrac{1}{2} \xi\phi^2$, 
respectively. Under a
conformal transformation $g_{\mu\nu} \rightarrow \tilde{g}_{\mu\nu} 
= \Omega^2 g_{\mu\nu}$, the action (\ref{FT:Snm}) transforms as follows:
\begin{eqnarray}
  S\rightarrow \tilde{S} =\int
  d^dx\,|g|^\frac{1}{2}\Omega^d\{\tilde{\cL}+\tfrac{1}{2}B\Omega^{-2}
  [R ~~~~~~~~~~~~~~~\,\nn \\-\,2(d-1)\Box\ln \Omega
  -(d-2)(d-1)(\nabla\ln \Omega)^2]\},
    \label{FT:Snm_tranf}
\end{eqnarray}
where $\tilde{\cL}$ is within the same class as $\cL$, and all indices
are raised and lowered with the untransformed metric.  Notice that if
we choose $\Omega = B^{1/(2-d)}$ and integrate
Eq. (\ref{FT:Snm_tranf}) by parts, the conformally transformed action
becomes the action for minimally coupled scalars with a Lagrangian
density of no larger generality than ${\cal L}$. (Derivatives
of $B$ give terms involving gradients of the scalar fields; after
adding these with additional multiplication by $\phi_a$-dependent
terms, the resulting Lagrangian is still within a class of ${\cal L}$
considered here.)  The conformally transformed action is
\begin{eqnarray}
  \tilde{S}=\int
    d^dx\,|g|^\frac{1}{2}[\tfrac{1}{2}R+B^{\frac{d}{2-d}}\tilde{\cL}
    -\tfrac{d-1}{2(d-2)}(\nabla \ln B)^2].
    \label{FT:Snm_final}
\end{eqnarray}
Thus, there is no need to consider non-minimally coupled models of 
Eq. (\ref{FT:Snm}) separately.

\begin{figure}
\includegraphics[width=8.5cm]{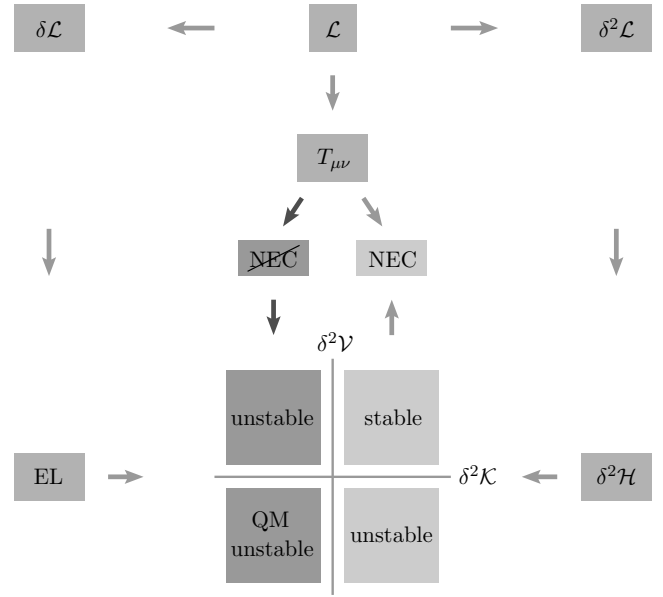}
\caption{A flow chart of the general version of the proof that for
models described by the action (\ref{FT:S}), only solutions satisfying
the NEC can be stable.}
\label{figure_chart}
\end{figure}

For the action
\begin{eqnarray}
  S=\int
  d^dx\,|g|^\frac{1}{2}\cL(\phi_a,D_\mu\phi_a,F_{a\mu\nu})\label{FT:S}
\end{eqnarray}
to be stationary, its first variation has to vanish, $\delta
S=0$. This leads to the equations of motion for the fields $\phi_a$
and $A_{a\alpha}$; in the classical analysis we assume that we have
found solutions to these equations, about which we expand. See Fig.
\ref{figure_chart} for a flow chart of the method that we use to show
a connection between violation of the NEC and instability.

\subsubsection{NEC}

We use the notation $\psi_{A\mu}=(D_\mu\phi_a,F_{a\alpha\mu})$, where
the abstract index $A$ runs over both Lorentz and color indices, as
well as type of field.  For example, in Sec. \ref{scalar}, the index
$A$ is a color index, $\psi_{A\mu}=\phi_{a,\mu}$, while in
Sec. \ref{gauge}, it is a Lorentz index, $\psi_{A\mu}=F_{\a\mu}$. In
general, for a model with $N_{\text{s}}$ scalar fields and
$N_{\text{g}}$ gauge fields in $d$ dimensions, $A$ runs from 1 to
$N_{\text{s}}+ \tfrac{1}{2} (d-1) N_{\text{g}}$.

\begin{figure}
\includegraphics[width=7.8cm]{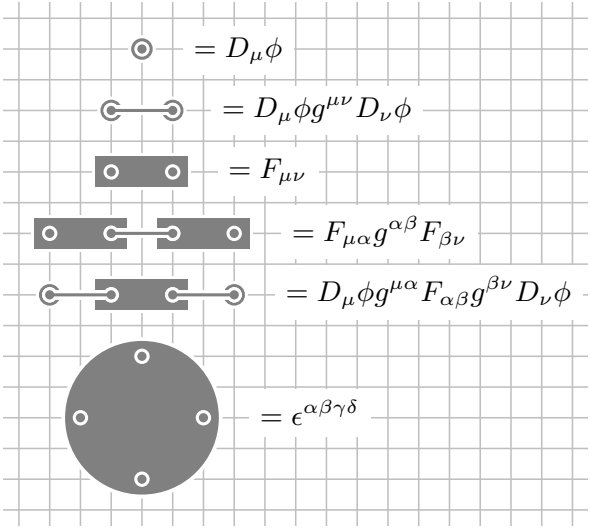}
\caption{Representation of simple Lorentz invariants
that the Lagrangian may depend upon. Each dot represents a Lorentz
index, and a line connecting them denotes contraction using the
metric. Rectangles (with two indices) are field strengths, small
circles covariant derivatives of scalar fields, and large circles
epsilon tensors. All Lorentz indices are ultimately contracted, and we
suppress color indices for simplicity.}
\label{figure_Lsimp}
\end{figure}

Assuming that $\psi_{A\mu}$ and $g^{\mu\nu}$ are independent
variables, we now prove that there is a relation between the
derivatives of $\cL$ with respect to them.  Since we consider only
Lorentz invariant models, the Lagrangian can be written in
terms of a number of Lorentz invariant quantities:
\begin{eqnarray}
  \cL=\cL(X_1,X_2,\ldots).
\end{eqnarray}
Objects represented in Fig.~\ref{figure_Lsimp} are the simplest
examples of $X_n$.  Each dot represents a Lorentz index, and a line
connecting them denotes contraction using the metric. Rectangles (with
two indices) are field strengths, small circles covariant derivatives
of scalar fields, and large circles epsilon tensors. All Lorentz
indices are ultimately contracted, and we suppress color indices for
simplicity.  As we will show, the relation between the derivatives of
$\cL$ is
\begin{eqnarray}
  \cL_{g^{\mu\nu}}&=&\sum_n \cL_{X_n}(X_n)_{g^{\mu\nu}} \nn\\ 
  &=& \tfrac{1}{2}\sum_n
    \left({M_{(n)}}^{AB}\psi_{A\mu}\psi_{B\nu} +g_{\mu\nu}K_{(n)}\right)\nn\\
  &=& \tfrac{1}{2}M^{AB}\psi_{A\mu}\psi_{B\nu} +\tfrac{1}{2}g_{\mu\nu}K ,
    \label{FT:M.g}\\
  \cL_{\psi_{A\mu}}&=&\sum_n \cL_{X_n}(X_n)_{\psi_{A\mu}} \nn\\
  &=&\sum_n \left({M_{(n)}}^{AB}{\psi_B}^\mu
  +\epsilon^{\mu\nu_2\cdots\nu_d}{{L_{(n)}}^A}_{\nu_2\cdots\nu_d}\right)\nn\\
  &=& M^{AB}{\psi_B}^\mu
  +\epsilon^{\mu\nu_2\cdots\nu_d}{L^A}_{\nu_2\cdots\nu_d}\label{FT:M.psi},
\end{eqnarray}
where quantities $M_{(n)}$, $K_{(n)}$ and $L_{(n)}$ originate from the
dependence of $\cL$ on $X_n$.  The first term in Eq.~(\ref{FT:M.g}) is
obtained by noting that for each and every $g^{\mu\nu}$ in $\cL$ there
are two $\psi$s attached to it. The second term is present if $\cL$
has a term involving the curved space totally antisymmetric tensor
$\vert g \vert^{-\frac{1}{2}} \epsilon^{\nu_1 \cdots \nu_d}$, again
recalling that the derivative of the determinant of the metric is
proportional to the metric.  Similarly, differentiation with respect
to $\psi_{A\mu}$ yields the $M$ and $L$ terms in Eq.~(\ref{FT:M.psi}).


\begin{figure}
\includegraphics*[width=6.5cm]{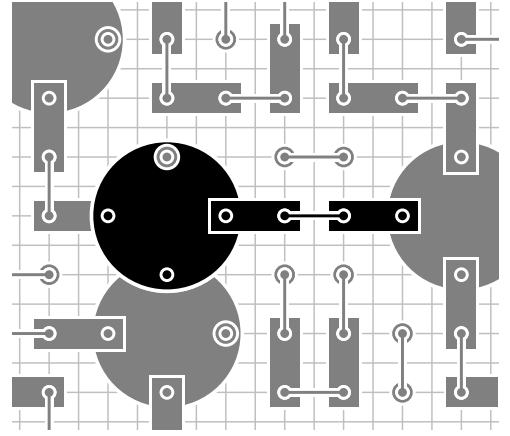}
\caption{Representation of the most general Lagrangian of the type
considered in this paper. The notation is the same as in
Fig.~\ref{figure_Lsimp}. We focus on the elements shaded black.  In
graphical terms, $M$ for this term in the Lagrangian can be obtained
from the figure by simply removing these elements. $Z$ represents the 
remainder of the Lagrangian.}
\label{figure_Lgen}
\end{figure}

To prove this relation for the most general case, we consider the most
general Lagrangian of the type considered in this paper, represented
by Fig.~\ref{figure_Lgen}.  The notation is the same as in
Fig.~\ref{figure_Lsimp}, with the addition that $Z$ denotes the
elements of the Lagrangian which are shaded gray. We focus our
attention on the term represented by the elements which are shaded
black.

Indeed, consider the portion of Fig.~\ref{figure_Lgen} whose elements
are shaded black; it equals
\begin{equation}
\cL = \vert g \vert^{-\frac{1}{2}} \epsilon^{\alpha \cdots} F_{\alpha
\mu} \, g^{\mu \nu} F_{\nu \beta} \, g^{\beta \gamma} Z_{\gamma
\cdots}.
\label{FT:M.L}
\end{equation}
(Again, we suppress color indices for simplicity, as they do not
affect the proof.) By differentiation, we obtain Eqs.~(\ref{FT:M.g})
and (\ref{FT:M.psi}) with
\begin{eqnarray}
M^{\alpha \beta} &=& - \vert g \vert^{-\frac{1}{2}} \left(
\epsilon^{\alpha \cdots} g^{\beta \gamma} + \epsilon^{\beta \cdots}
g^{\alpha \gamma} \right) Z_{\gamma \cdots}, \\ K &=& \vert g
\vert^{-\frac{1}{2}} \epsilon^{\alpha \cdots} F_{\alpha \rho} \,
g^{\rho \sigma} F_{\sigma \beta} \, g^{\beta \gamma} Z_{\gamma \cdots}, \\
{L^\alpha}_{\cdots} &=& - \vert g \vert^{-\frac{1}{2}} g^{\alpha \rho}
F_{\rho \beta} \,g^{\beta \gamma} Z_{\gamma \cdots}. \label{FT:L}
\end{eqnarray}
Note the derivative $\cL_{F_{\nu \beta}}$ generates a contribution to
$M$ which is matched by a corresponding contribution from
$\cL_{g^{\beta \gamma}}$. Other contractions of fields with $g_{\mu
\nu}$ (i.e., as indicated in the figure) can be analyzed similarly.
In graphical terms, $M$ can be obtained from the figure by simply
removing the elements shaded black.

From Eq.~(\ref{FT:M.g}), the NEC requires
\begin{eqnarray}
  \Psi_A M^{AB}\Psi_B\ge 0, \label{FT:NEC.M}
\end{eqnarray}
where $\Psi_A=\psi_{A\mu}n^\mu$. Thus, to satisfy the NEC, $M^{AB}$ has to
be positive semidefinite on the subspace spanned by $\Psi_A$. This 
property is crucial for stability of solutions, to which we now turn.

\subsubsection{Stability}

To study the stability of the solution $\psi_A(x)$, we consider the
second variation of the action,
\begin{eqnarray}
  \d^2 S = \int d^dx\,|g|^\frac{1}{2}\left(\cL_{\psi_A\psi_B}
  \delta\psi_A\delta\psi_B\right.\nn\\ + \left.
  2\cL_{\psi_{A}\psi_{B;\lambda}}\delta\psi_{A}\delta\psi_{B;\lambda}
  + \cL_{\psi_{A;\mu}\psi_{B;\nu}}
  \delta\psi_{A;\mu}\delta\psi_{B;\nu}\right).\label{FT:d2S}
\end{eqnarray}
Here quantities $\cL_{\psi_A}=\p\cL/\p\psi_A$, etc. are evaluated at
$\psi_A(x)$. Also notice that
$\psi_{A;\mu}=(D_\mu\phi_a,A_{a\alpha;\mu})$, the covariant
derivatives of $\psi_A$, are different from
$\psi_{A\mu}=(D_\mu\phi_a,F_{a\alpha\mu})$.

Differentiating Eq.~(\ref{FT:M.psi}) we find
\begin{equation}
\cL_{\psi_{A\mu}\psi_{B\nu}} = M^{AB}g^{\mu\nu} +N^{A\mu
B\nu}.\label{FT:MN}
\end{equation}
The separation of the second derivative into the first and second
terms in Eq.~(\ref{FT:MN}) is natural: $g^{\mu\nu}$ appears only in
the first term, and $N$ represents all remaining terms. $N$ contains
terms obtained by differentiating $M$ and $L$ with respect to
$\psi_{B\nu}$, plus additional terms if $\psi$ is a field strength, as
in Eq.~(\ref{A:N}). The $\nu$ index obtained from these $\psi_{B\nu}$
derivatives is attached to a field and not the metric $g^{\mu
\nu}$. ($L$ does not contain an epsilon tensor since $Z$ does not.)

Next notice that even though $\psi_{A\mu}$ and $\psi_{A;\mu}$ differ,
the derivatives of $\cL$ with respect to them are related due to the
form of the action~(\ref{FT:S}). Thus, the second variation of the
Lagrangian in Eq.~(\ref{FT:d2S}) becomes
\begin{eqnarray}
  \delta^2\cL &=&\left(M^{AB}g^{\mu\nu} +N^{A\mu B\nu}
  \right)\delta\psi_{A;\mu}\delta\psi_{B;\nu}\nn\\
  &+&2\cL_{\psi_{A}\psi_{B;\lambda}}
  \delta\psi_{A}\delta\psi_{B;\lambda} -\d^2 V, \label{FT:d2L}
\end{eqnarray}
where $\d^2 V= -\cL_{\psi_A\psi_B}\d\psi_A\d\psi_B$.

Let us use a locally inertial frame. For the
Lagrangian~(\ref{FT:d2L}), the canonical momentum is
\begin{eqnarray}
  \delta{\pi}^A =2{M}^{AB}\delta{\psi}_{B;0}
  +2{N}^{A0B\nu}\delta{\psi}_{B;\nu}\nn\\
  +2\cL_{\psi_{B}\psi_{A;0}}\delta\psi_{B},
  \label{FT:dpi}
\end{eqnarray}
which leads to the following effective Hamiltonian for fluctuations
about the classical solution:
\begin{eqnarray}
\delta^2\cH
    =\delta^2 K +\d^2 V -2\cL_{\psi_{A}\psi_{B;i}}
    \d\psi_{A}\d\psi_{B;i},\label{FT:d2H}
\end{eqnarray}
where the kinetic term is
\begin{eqnarray}
\delta^2 K &=& \left({M}^{AB}+{N}^{A0B0}\right)
    \delta\psi_{A;0}\delta\psi_{B;0}\nn\\ &+&
    \left({M}^{AB}\delta^{ij} -{N}^{AiBj}\right)
    \delta\psi_{A;i}\delta\psi_{B;j}.\label{FT:d2K}
\end{eqnarray}
Here $\delta{\psi}_{A;0}$ are functions of $\delta{\pi}_B$, $\d
{\psi}_B$, and $\delta{\psi}_{B;i}$ as found from Eq.~(\ref{FT:dpi}).
 
We now prove that nonnegativeness of the kinetic term $\delta^2 K$
implies positive semidefiniteness of the matrix $M^{AB}$. Indeed,
suppose that $M^{AB}$ is not positive semidefinite. In such a case,
the matrix $M^{AB}$ has at least one negative eigenvalue, which means
that there is a basis in which the matrix is diagonal with at least
one negative entry, ${M}^{AB}=\text{diag\,}({m}_1,\ldots,{m}_n)$
(${m}_1<0$). Let us choose field variations which are nonzero only in
the direction of the negative eigenvalue: $\delta{\psi}_{1;\mu}\not
=0$ and $\delta{\psi}_{A;\mu}=0$ $(A>1)$. We further restrict $d-1$
quantities $\delta{\psi}_{1;i}$ to satisfy the following equation:
\begin{eqnarray}
  {N}^{1010}\delta{\psi}_{1;0} \delta{\psi}_{1;0}
  ={N}^{1i1j}\delta{\psi}_{1;i}\delta{\psi}_{1;j}.
  \label{FT:Ncondition}
\end{eqnarray}
These conditions make the kinetic term in Eq.~(\ref{FT:d2H}) negative,
\begin{eqnarray}
  \delta^2 K ={m}_1\left[(\delta{\psi}_{1;0})^2
    +\sum_i(\delta{\psi}_{1;i})^2\right] <0,
  \label{FT:d2K1}
\end{eqnarray}
thus proving that in order for $\delta^2 K$ to be nonnegative, the
matrix $M^{AB}$ has to be positive semidefinite.

Note that if the background field leaves an unbroken rotational
symmetry---i.e., there are $d-2$ or fewer independent vacuum
expectation values $\psi_\mu$---one can find a solution to
Eq.~(\ref{FT:Ncondition}) in {\it any} frame, by taking
$\delta{\psi}_{1;0} = 0$ and the spacelike fluctuations to have
support only in the direction(s) orthogonal to all of the
$\psi_\mu$. In the remaining cases (which, in particular, have no
unbroken rotational symmetries) there may be Lorentz frames in which
no solution to Eq.~(\ref{FT:Ncondition}) exists. The examples given by
Ref.~\cite{dubovsky} are in this category.  In Refs.~\cite{dubovsky}
and \cite{nima}, it is emphasized that superluminal modes play a
crucial role in permitting stability and NEC violation
simultaneously. However, models with superluminal modes are not
causal. In the appendix, we show that the assumption of causality is
strong enough to restore the direct connection between NEC violation
and instability.

If the kinetic energy is negative, then the system described by the
Hamiltonian (\ref{FT:d2H}) is (locally) unstable. The linear term in
Eq.~(\ref{FT:d2H}) can never act to stabilize the system. Therefore,
in order to determine whether the system is unstable, we focus only on
the quadratic terms.  If the potential energy is positive then small
perturbations will cause the classical solutions to grow exponentially
away from the original stationary point. However, it is possible to
have classical stability if one chooses the potential energy to be
negative; in this case we have an upside-down potential with negative
kinetic term, or a ``phantom''. Such models necessarily exhibit
quantum instabilities \cite{phantom}.  Using the result established in
the previous paragraph, we thus conclude that solutions to the theory
given by the action~(\ref{FT:S}) are stable only if the matrix
$M^{AB}$ is positive semidefinite.
 
Combining the relations between nonnegativeness of $\delta^2 K$ and
positive semidefiniteness of $M^{AB}$ on one hand, and the NEC and
positive semidefiniteness of $M^{AB}$ on the the other hand, we
conclude that for the theory given by the action~(\ref{FT:S}), only
solutions satisfying the NEC can be stable.

\section{Quantum field theories} \label{quantum}

We can deduce similar results for quantum systems.  Suppose there
exists a quantum state $\vert \alpha \rangle$ and a null vector
$n^\mu$ such that the NEC is violated in a quantum averaged sense:
\begin{eqnarray}
  \langle \alpha \vert T_{\mu\nu} \vert \alpha \rangle n^{\mu} n^{\nu}
  =\langle \alpha \vert M^{AB} \Psi_{A} \Psi_{B} \vert \alpha \rangle
  <0.
\end{eqnarray}
Define a
basis $\vert \phi \rangle$ in which the operator ${\cal M} = M^{AB}
\Psi_A \Psi_B$ is diagonal:
\begin{equation}
{\cal M} \vert \phi \rangle = m(\phi) \vert \phi \rangle.
\end{equation}
By orthonormality of $\vert \phi \rangle$, we see that violation of
the quantum averaged NEC implies
\begin{eqnarray}
  \langle \alpha \vert {\cal M} \vert \alpha \rangle &=&\sum_{\phi
  \phi'} \, \langle \alpha \vert \phi \rangle \langle \phi \vert {\cal
  M} \vert \phi' \rangle \langle \phi' \vert \alpha \rangle\nn\\ &=&
  \sum_{\phi} \, \vert \langle \alpha \vert \phi \rangle \vert^2
  \,m(\phi) < 0.
\end{eqnarray}
This means that there exist eigenstates $\vert \phi \rangle$, whose
overlap with $| \alpha \rangle$ is non-zero and on which the operator
${\cal M}$ has negative eigenvalues. This requires that $M^{AB}$ and
hence $\delta^2 K$ is not positive semidefinite; by continuity, this
must also be the case in a ball $B$ in the Hilbert space of $\vert
\phi \rangle$.

As a further consequence, we can conclude that a state $\vert \alpha
\rangle$ in which the quantum averaged NEC is violated cannot be the
ground state~\cite{OW}. Suppose that $\vert \alpha \rangle$ is an energy eigenstate:
$H \vert \alpha \rangle = E_\alpha \vert \alpha \rangle$. An
elementary result from quantum mechanics is that $\vert \alpha
\rangle$ can be the ground state only if
\begin{equation}
E_\alpha = \langle \alpha \vert H \vert \alpha \rangle \leq \langle
\alpha' \vert H \vert \alpha' \rangle
\end{equation} 
for all normalized states $\vert \alpha' \rangle$ which need not be
energy eigenstates. However, it is possible to reduce the expectation
value of $H$ by perturbing $\vert \alpha \rangle$. Specifically, we
adjust $\vert \alpha \rangle$ only in the ball $B$, where we know from
Eqs.~(\ref{FT:d2K})--(\ref{FT:d2K1}) that there are perturbations
which reduce the expectation of the kinetic energy without changing
the expectation of the potential. For a pictorial representation of
this, see Fig.~\ref{figure_grad}, where now the two different graphs
are sketches of two eigenstates of the operator $\cM$. Let the top one
be $\vert \phi_1 \rangle$ and the bottom $\vert \phi_2 \rangle$.
Expand the state $\vert \alpha \rangle$ in the basis of these
eigenstates:
\begin{eqnarray}
  \vert \alpha \rangle = \sum_i c_i \vert \phi_i \rangle .
\end{eqnarray}
Since by assumption the state $\vert\alpha\rangle$ violates the
quantum averaged NEC, the magnitude of coefficients of smooth
eigenstates, e.g. $\vert \phi_1 \rangle$, will decrease, while those
of eigenstates with large gradients, e.g.  $\vert \phi_2 \rangle$,
will increase, until the quantum averaged NEC is no longer
violated. Thus, a state $\vert\alpha\rangle$ which violates the
quantum averaged NEC cannot be the ground state because perturbations
within the ball $B$ can lower the expectation value of the energy.

Note that the discussion above is in
terms of unrenormalized (bare) quantities. The renormalized
expectation 
\begin{eqnarray}
  \langle \alpha \vert {\cal M}_{\rm ren} \vert \alpha\rangle =
    \langle \alpha \vert {\cal M} \vert \alpha \rangle - \langle 0
    \vert {\cal M} \vert 0 \rangle,
\end{eqnarray}
where $\vert 0 \rangle$ is the flat space QFT ground state, could be
negative (e.g., as in the Casimir effect \cite{null}), but this is
possible only if $\vert \alpha \rangle$ is not the ground state.

In known cases of NEC violation, such as the Casimir vacuum or black
hole spacetime, it is only the {\it renormalized} energy-momentum
tensor which violates the NEC. As a simple example, consider a real
scalar field $\phi$. The energy-momentum tensor is simply $T_{\mu \nu}
= \partial_\mu \phi \, \partial_\nu \phi$ plus terms proportional to
$g_{\mu\nu}$ which do not play a role in the NEC. Then ${\cal M} = (
n^{\mu} \partial_{\mu} \phi )^2$ is a Hermitian operator with positive
eigenvalues. Therefore, its expectation value in {\it any} state is
positive:
\begin{eqnarray}
  \langle \alpha \vert {\cal M} \vert \alpha \rangle  >  0,
\end{eqnarray}
for any $\vert \alpha \rangle$, including the Hartle-Hawking, Casimir
or flat-space vacuum. We can verify this by direct calculation in
$d=4$ dimensions, computing the energy-momentum tensor using
point-splitting regularization:
\begin{eqnarray}
  \langle 0 \vert \phi(x) \phi(x') \vert 0 \rangle = -\frac{1}{4 \pi^2
  \vert x-x' \vert^2},
\end{eqnarray}
which yields
\begin{eqnarray}
  \langle 0 \vert T_{\mu\nu} (x,x') \vert 0 \rangle n^\mu n^\nu =
  \frac{2 [n^\mu (x-x')_\mu]^2 }{\pi^2 \vert x-x'\vert^6}.
\end{eqnarray}
So $\langle 0 \vert {\cal M} \vert 0 \rangle$ is manifestly
positive. Note that this result holds even if the expectation is taken
in an arbitrary state rather than $\vert 0 \rangle$, since the bare
expectation is always dominated by the UV contribution. Now, had we
taken a {\it negative} kinetic energy term for the scalar, the overall
sign of ${\cal M}$ changes, allowing violation of the NEC. But, this
model is clearly unstable, in accordance with our results.

Our results are summarized as follows, where $\vert \alpha \rangle$ is
any quantum state, and only models of the type described by action
(\ref{FT:S}) are considered: (1) $\langle \a \vert {\cal M} \vert \a
\rangle < 0$, i.e., the bare NEC is violated, implies that $\vert \a
\rangle$ is locally unstable, (2) $\langle \a \vert {\cal M}_{\rm ren}
\vert \a \rangle < 0$, i.e., the renormalized NEC is violated, implies
that $\vert \alpha \rangle$ cannot be the global ground state.

\section{Models with fermions} \label{fermions}

In this section we present a generalized version of the treatment of
fermions in \cite{buniy}.  Both Dirac and Weyl fermions with standard
couplings to the bosonic fields of Sec. \ref{quantum} may be included
in the model without any effect on the results of that section.  In
the purely classical analysis, fermions play no role, while in the
quantum case we can integrate them out to obtain their contribution to
the energy-momentum tensor.  We will see that this fermionic
contribution always satisfies the NEC.

In addition to the bosonic part in the Lagrangian, we include the
following fermionic part:
\begin{eqnarray} 
\cL^{(\text{f})}=\bar\psi\left[i\D-m(\phi_a)\right]\psi.\label{cL.fermions}
\end{eqnarray}
A Dirac fermion $\psi$ couples to scalar and vector bosons via a
covariant derivative $D_\mu=\nabla_\mu+A_\mu$ and an arbitrary real
scalar function $m(\phi_a)$. We use Euclidean quantities obtained via
substitutions $x^0 \to -ix^0$, $A_0 \to iA_0$, $\gamma^0 \to
-i\gamma^0$, and find the effective Euclidean action $\Gamma$ from
\begin{eqnarray} 
e^{-\Gamma}&=&\int\cD\bar\psi\cD\psi\exp{\left[-\int d^dx \vert
g\vert^\frac{1}{2} \,\cL^{(\text{f})}\right]} \nn \\ &=&\text{det\,}
(i\D-m).\label{Gamma}
\end{eqnarray}
The Euclidean operator $\D$ is self-adjoint.

The imaginary part of the action vanishes since it is an odd function
of $\D$ and the trace of odd number of gamma matrices vanishes:
\begin{eqnarray} 
\Im\Gamma&=&-\tfrac{1}{2}\tr\ln\left[(i\D-m)(-i\D-m)^{-1}\right] \nn
  \\ &=&i\,\tr\text{arctan\,}(\D m^{-1})=0.
\end{eqnarray}
Using eigenvectors $\psi_k$ and eigenvalues $\lambda_k$ of the
operator $i\D-m$, the real part of the action becomes
\begin{eqnarray} 
\Re\Gamma &=&-\tfrac{1}{2}\tr\ln{[(-i\D-m)(i\D-m)]} \nn \\
  &=&-\tfrac{1}{2}\sum_k\ln{\vert\lambda_k\vert^2}.
\label{Gamma.Re}
\end{eqnarray}
Since the operator $i\D-m$ is not self-adjoint, the eigenvalues are
complex and the eigenfunctions are not orthogonal. However, they can
be normalized, $\int d^dx \vert g\vert^\frac{1}{2}
\psi^\dagger_k\psi_k=1$. Using this normalization we have
\begin{eqnarray} 
\vert\lambda_k\vert^2 =\int d^dx \vert g\vert^\frac{1}{2}
 \psi^\dagger_k(-i\D-m)(i\D-m)\psi_k.
\label{Lambda2}
\end{eqnarray}
For the purposes of calculating the contribution of $\text{Re\,}\Gamma$ to
the energy momentum tensor, it is enough to consider only those terms
in Eq.~(\ref{Lambda2}) which contain $g^{\mu\nu}$:
\begin{eqnarray} 
\vert\lambda_k\vert^2 =-\int d^dx \vert g\vert^\frac{1}{2}
g^{\mu\nu}(D_\mu\psi_k)^\dagger (D_\nu \psi_k)+\cdots ~.
\label{Lambda3}
\end{eqnarray}
We have used here anti-hermiticity of $D_\mu$. Thus, using the
definition of the energy momentum tensor
\begin{eqnarray}
T^{\text{(f)}}_{\mu\nu} = 2\vert g\vert^{-\frac{1}{2}}
    \frac{\d \Gamma}{\d g^{\mu\nu}},
\end{eqnarray}
the fermionic
contribution to the NEC expression is non-negative:
\begin{eqnarray}
T^{\text{(f)}}_{\mu\nu}n^\mu n^\nu
=\sum_k\vert\lambda_k\vert^{-2}(n^\mu D_\mu\psi_k)^\dagger (n^\nu
D_\nu \psi_k)\ge 0.
\label{T.fermion}
\end{eqnarray}

The Weyl operator does not define an eigenvalue problem, as it maps a
spinor into the opposite chirality; see e.g.~\cite{Weyl}. However, the
Weyl determinant can be defined as the square root of the
corresponding Dirac determinant obtained by considering the Weyl
fermion and a partner in the complex conjugate representation. The
real part of the contribution to the energy-momentum tensor is exactly
one half of the Dirac contribution considered previously.  The
presence of either type of fermion does not lead to violation of the
NEC, and so the results of Sec.~\ref{quantum} hold in models with
arbitrary fermionic and bosonic particle content.

\section{Fluids} \label{fluids}

\subsection{Perfect Fluids}

A macroscopic system may be approximately described as a perfect fluid
if the mean free path of its components is small compared to the
length scale of interest. For the dark energy, this length scale is of
cosmological size.  A perfect fluid is described by the
energy-momentum tensor
\begin{eqnarray}
T_{\mu\nu}=(\rho+p)u_\mu u_\nu-pg_{\mu\nu},\label{F:T}
\end{eqnarray}
where $\rho$ and $p$ are the energy density and pressure of the fluid
in its rest frame, and $u_\mu$ is its velocity. Let $j^\mu=J u^\mu$ be
the conserved current vector (${j^\mu}_{;\mu}=0$), and $J=(j_\mu
j^\mu)^\frac{1}{2}$ the particle density.

The energy-momentum can be written \cite{Hawking-Ellis,Jackiw:2004nm}
as
\begin{eqnarray}
T_{\mu\nu}=(f-J f')g_{\mu\nu}+(f'/J) j_\mu j_\nu,\label{F:T1}
\end{eqnarray}
where, comparing with Eq.~(\ref{F:T}), we have $\rho = f(J)$ and $p =
Jf' - f$. The function $f(J)$ implicitly determines the equation of
state.  Also, prime indicates derivative with respect to $J$ at fixed
temperature and volume.
 
The NEC for the tensor~(\ref{F:T1}) becomes
\begin{eqnarray}
T_{\mu\nu}n^\mu n^\nu=(f'/J)(j_\mu n^\mu)^2\ge 0.\label{F:NEC}
\end{eqnarray}
Thus, perfect fluids with negative $f'(J)$ violate the NEC. Below, we
demonstrate that $f'(J) < 0$ implies an instability.

Recall that the speed of sound in a fluid is given by $c_\text{s} =
(dp / d\rho)^{\frac{1}{2}} = (J f'' / f')^{\frac{1}{2}}$, and that
complex $c_\text{s}$ (or negative compressibility) implies a
mechanical instability \cite{neg.comp.}. This instability is avoided
only if $f'' < 0$. However, if $f'$ and $f''$
are negative, then the fluid is unstable with respect to
clumping. To see this, we first recall that the Helmholtz free energy
$F$ is a Legendre transform of the internal energy $E$, i.e.,
$F=E-TS$, where $S$ is the entropy. In the thermodynamic limit,
entropy scales with particle number $N=JV$, and so
\begin{eqnarray}
F = Vf - JVTs,
\end{eqnarray}
where from above, $f(J)=\rho$ is the energy density, and 
$s=S/N$ is the specific entropy, which is a function
of temperature but not particle number. Therefore,
\begin{eqnarray}
  \left.\frac{\p^2 F}{\p J^2}\right|_{T,V} = Vf''.\label{fluid:f''}
\end{eqnarray}

Now consider two adjacent regions of the fluid with identical
volumes. Suppose we transfer a small amount of matter $\delta J$ from
one volume to another. Expand the change in free energy to second
order in the number density:
\begin{eqnarray}
  \d F_1 + \d F_2 &=& \frac{\p F}{\p J_1} \d J_1 + \frac{1}{2}\frac{\p^2 F}{\p
    J_1^2} (\d J_1)^2 \nn\\ &+& \frac{\p F}{\p J_2} \d J_2 +
    \frac{1}{2}\frac{\p^2 F}{\p J_2^2} (\d J_2)^2 + \cdots ~.
\end{eqnarray}
The linear terms cancel due to particle number conservation, $\d J_1 =
- \d J_2 = \d J$. Therefore, using Eq.~(\ref{fluid:f''}), the
resulting change in total free energy is given by
\begin{eqnarray}
  \d F = \tfrac{1}{2} V f''(J) (\delta J)^2 <0.
\end{eqnarray}
We see that the system can decrease its free energy by clumping into
over- and under-dense regions. This itself is an instability, which
results in a runaway to infinitely negative free energy unless the
assumption of negative $f'$ or negative $f''$ ceases to hold. 
NEC violation either leads complex speed of sound or clumping instability.

\subsection{Beyond Fluids}

We can generalize our analysis beyond fluids, relaxing the assumption
of the existence of a conserved current $j^\mu$, and requiring only
energy conservation and statistical equilibrium.

Consider an isolated system at fixed energy $E$. If the system is
ergodic it can (after a sufficiently long time) be described by a
microcanonical ensemble, which uniformly samples phase space. If it
has many degrees of freedom, the microcanonical ensemble is well
approximated by a canonical ensemble with a Boltzmann weighting, in
which the energy is allowed to fluctuate. In this case the
thermodynamical relation $E = TS - pV$ is satisfied, with $T =
(\partial E / \partial S) \vert_V$.  Dividing by the volume, we obtain
$\rho + p = T \sigma$, where $\sigma = S/V$ is the positive-definite
entropy density.  Thus, for a system which is homogeneous and
isotropic, i.e., one with energy-momentum tensor as in
Eq.~(\ref{F:T}), violation of the NEC requires negative temperature,
or entropy (number of states) which decreases with energy.  For a
field theory, this suggests either a negative kinetic energy term or
an ``upside-down'' potential; if the kinetic and potential energies
have opposite sign (the usual case), the number of states available to
the system increases as energy is added, and the temperature is
positive.  Moreover, in order for the partition function (trace over
Boltzmann weights) to converge, the kinetic energy term must be
negative if the temperature is. This is because at any fixed potential
energy there is an infinite volume of phase space with arbitrarily
large gradients, i.e., kinetic energy. Therefore, if the kinetic
energy term and the temperature are not of the same sign, the
partition function does not converge.

An example of a condensed matter system with a negative temperature is
a ferromagnet in an external magnetic field \cite{neg.temp}.  Beyond a
certain threshold energy, the number of states available to the system
decreases as energy is added, i.e., the temperature is negative. Due
to the presence of an external magnetic field, however, the system is
anisotropic; its energy-momentum tensor is not of the form of
Eq.~(\ref{F:T}), and there is no connection between negative
temperature and violation of the NEC. It is only for homogeneous,
isotropic systems, e.g., systems of interest in cosmology, that this
relation exists.

\section{Conclusion} \label{conclusion}

We have exhibited a direct connection between violation of the NEC and
instability in a wide variety of models. One important
application is to the dark energy, whose equation of state $w$ should
be no more negative than $-1$ in order not to violate the NEC.

In classical field
theory---including all Lorentz-invariant models involving both
minimally and non-minimally coupled scalar and gauge fields with
second-order equations of motion---violation of the NEC generally
implies the existence of an instability toward the formation of
gradients. In the case of configurations that possess at least some
rotational symmetry (e.g., those of interest in cosmology), violation
of the NEC always implies instability (Sec. IIC). In cases with less
symmetry, it is possible for NEC-violating models to be stable if they
exhibit superluminal excitations \cite{dubovsky}; however, such models 
are not causal \cite{nima}. In the appendix we show that in any causal
scalar model, NEC violation implies instability.

These results can be generalized to quantum theories, including
fermions. Analogous results for perfect fluids relate violation of the
NEC to either a complex speed of sound or a clumping instability.

\begin{acknowledgments}

The authors thank Alejandro Jenkins for very useful comments. 
This work was supported by the Department of Energy under
DE-FG06-85ER40224.

\end{acknowledgments}

\appendix*

\section{Causality and stability}

In this appendix we impose an additional causality constraint, which
excludes superluminal excitations of the type considered in
Refs.~\cite{dubovsky}, \cite{nima}. We show that a causal,
NEC-violating model cannot be stable. For simplicity, we limit our
attention to the case of scalar fields.

\subsection{Fluctuations}

Consider a theory of $n$ scalar fields $\phi_a$ in $d$-dimensional
spacetime. Using the first derivative of the field,
$\phi_{a\mu}=\nabla_\mu\phi_a$, we can construct a Lorentz-invariant
matrix $X_{ab}=g^{\mu\nu}\phi_{a\mu}\phi_{b\nu}$. The most general
Lorentz invariant Lagrangian resulting in the equation of motion of at
most second order is $\cL=F(X_{ab},\phi_c)$, where the function $F$ is
a Lorentz scalar, but in general it is not a scalar in the flavor
space. The second fluctuation of the Lagrangian is
\begin{eqnarray} 
  \delta^2\cL=F^{a\mu b\nu}\delta\phi_{a\mu}\delta\phi_{b\nu}
  +\cdots,\label{d2cL}
\end{eqnarray}
where 
\begin{eqnarray} 
  F^{a\mu b\nu}=F_{X_{ab}}g^{\mu\nu}
  +2F_{X_{ac}X_{bd}}\phi_c^\mu\phi_d^\nu.\label{F}
\end{eqnarray}
The equation of motion for the fluctuations is
\begin{eqnarray} 
  F^{a\mu b\nu}\delta\phi_{b\mu\nu}+\cdots=0.\label{fluctuations}
\end{eqnarray}
In Eq.~(\ref{d2cL}), ellipses denote terms at most linear in the
derivatives of fluctuations; such terms are not important for the
present analysis.  Specifically, they do not enter into the equation
for the characteristic surfaces or the dispersion relation for short
wavelength fluctuations. See, e.g., Ref.~\cite{petrovskii} for more on
characteristics in partial differential equations. In the case of the
dispersion relation, the short wavelength approximation dictates that
only terms with the largest number of derivatives matter.

\subsection{Hyperbolicity and causality}

When the equation of motion is hyperbolic, its domain of influence is
bounded by the outside characteristics originating from the region where
nontrivial initial data is prescribed.  The resulting characteristic
surface is given by $\xi(x)=0$, where $\xi_\mu=\nabla_\mu \xi$
satisfies the characteristic equation
\begin{eqnarray} 
  \left\|F^{a\mu b\nu}\xi_\mu
  \xi_\nu\right\|=0,\label{characteristics}
\end{eqnarray}
and $\left\| \cdots \right\|$ denotes a determinant in flavor space.
For a hyperbolic model, characteristics
and therefore $\xi_\mu$ are real. For such a model, the characteristic
surface is inside the light cone $\xi_0^2-\delta^{ij}\xi_i\xi_j<0$
(i.e., the model is causal), if and only if the matrix $F^{a\mu
b\nu}n_\mu n_\nu$ is sign-definite for all null $n_\mu$. By writing
$F^{a\mu b\nu} = \sum_c F_c^{\mu\nu} v^a_c v^b_c$, where
$F_c^{\mu\nu}$ and $v_c^a$ are eigenvalues and eigenvectors,
respectively, this result is obtained by noticing that the
characteristic surface consists of the null surfaces for the metrics
$F_c^{\mu\nu}$. Each of these null surfaces has to be inside the light
cone, hence the above result. From Eq.~(\ref{F}), this means that the
matrix $C^{ab}=2F_{X_{ac}X_{bd}}(\phi_c^\mu n_\mu)(\phi_d^\nu n_\nu)$
is sign-definite.

In terms of $A^{ab}=F_{X_{ab}}$,
$B^{ab}=2F_{X_{ac}X_{bd}}(\phi_c^\mu\xi_\mu)(\phi_d^\nu\xi_\nu)$, and
$\omega=-\xi_\mu\xi^\mu$, the characteristic equation is
\begin{eqnarray}
  \left\|B-\omega A\right\|=0.
\end{eqnarray}
For nonsingular $A$, the parameter $\omega$ is equal to one of the
eigenvalues of the matrix $A^{-1}B$. Causality requires $\omega>0$ and
so\,\footnote{The inequality $Q>0$ ($Q<0$) for the matrix $Q$ means
that $Q$ is positive (negative) definite.}  $A^{-1}B>0$.

Consider a Lorentz boost  
\begin{equation}
\phi_a^\mu\xi_\mu ~\rightarrow~ \tilde\phi_a^\mu \tilde\xi_\mu
\approx \tilde\phi_a^\mu n_\mu
\end{equation}
for some particular null vector $n_\mu$. (It is easy to check that
$\tilde\xi_\mu$ can always be made nearly null by appropriate choice
of frame.) In this new frame, $B^{ab}$ is arbitrarily close to the
matrix $C^{ab}$ evaluated on this particular $n_\mu$. Then, because
$C$ is sign-definite for all $n_\mu$, the sign-definiteness of $B$
follows. This means that if a model is causal, then either
$A>0$, $B>0$ or $A<0$, $B<0$. This result can be extended to singular
$A$ as well.

\subsection{Stability and NEC}

To study local stability, it is enough to consider the equation of
motion in a small neighborhood of a given point $x$. In such a
neighborhood, we can ignore spacetime dependence of $F^{a\mu b\nu}$
and solve the equation in terms of Fourier modes with momenta
$k_\mu$. Also, in the short wavelength approximation, only terms
quadratic in derivatives matter. The corresponding dispersion relation is
\begin{eqnarray} 
  \left\|F^{a\mu b\nu}k_\mu k_\nu\right\|=0.\label{dispersion}
\end{eqnarray}
Classical stability requires that $k_\mu$ is real, which means that
the matrix $F^{a\mu b\nu}$ satisfies the same condition which is
required for the characteristics to be real. In other words, models
are classically stable if and only if they are hyperbolic.  To avoid
ghosts, i.e, quantum mechanical instabilities \cite{phantom}, there is
an additional condition that the matrix $F^{a0b0}$ be positive
definite. Except in the trivial case where $F_{X_{ab}X_{cd}}$
vanishes, we can always choose such a Lorentz boost that $F^{a0b0}$ is
arbitrarily close to $C^{ab}$ for some $n_\mu$. Since $B$ and $C$ are
of the same sign definiteness, the matrix $F^{a0b0}$ can be positive
definite, and therefore the model is ghost-free, only if $B>0$. Then,
from the previous section, the combined requirement that the model be
both causal and ghost-free can be satisfied only if $A>0$ and $B>0$.

\begin{figure}
\includegraphics[width=6cm]{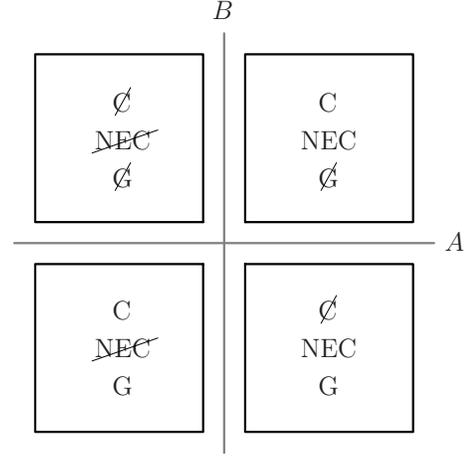}
\caption{The NEC, causality (here C), and quantum mechanical
stability (the absence of ghosts, here $\not\!\!{\rm G}$) are all
determined by the sign-definiteness of matrices $A$ and $B$. The NEC
holds for $A>0$; the model is causal for $A^{-1}B>0$ and ghost-free
for $B>0$. A model is ghost-free and causal only if it obeys the NEC.}
\label{appendix_fig}
\end{figure}

From the energy-momentum tensor
\begin{eqnarray} 
  T_{\mu\nu}=-Fg_{\mu\nu}+2F_{X_{ab}}\phi_{a\mu}\phi_{b\nu},
\end{eqnarray}
it follows that the NEC is equivalent to $A\ge 0$. Thus, for a
hyperbolic, causal ($A^{-1}B>0$), classically and quantum mechanically
stable model ($B>0$), the NEC is satisfied ($A\ge 0$). Vice versa, a
causal, hyperbolic (and therefore classically stable), NEC-violating
model cannot be quantum mechanically stable. See
Fig.~\ref{appendix_fig} for a diagram summarizing the relation between
quantum mechanical stability, causality, and the NEC.

\subsection{Example}

For the case $n=1$, combining Eqs.~(\ref{F}) and (\ref{characteristics}),
the characteristic equation is
\begin{eqnarray} 
  F_X\xi_\mu\xi^\mu+2F_{XX}(\phi^\mu\xi_\mu)^2=0,
\end{eqnarray}
and the model is causal if $F_X^{-1}F_{XX} > 0$. Violation of the NEC
and absence of ghost instability imply $F_X < 0$ and $F_{XX} > 0$,
respectively. We see that NEC violation and the absence of ghost
instability imply non-causal behavior for this model. Moreover, for
$F_X < 0$ and $F_{XX} > 0$ it is impossible to find a Lorentz frame in
which the Hamiltonian
\begin{eqnarray} 
  \delta^2\tilde\cH&=&\left[F_X+2F_{XX}(\tilde\phi^0)^2\right]
  (\delta\tilde\phi_0)^2 \nn \\
  &+&\left[F_X\delta^{ij}-2F_{XX}\tilde\phi^i\tilde\phi^j\right]
  \delta\tilde\phi_i\delta\tilde\phi_j +\cdots
\end{eqnarray}
is positive definite since the second term is always
negative. Clearly, this can happen only for a non-causal model. When
superluminal modes are present (i.e., a model is not causal), the sign
of the Hamiltonian can be changed by performing a Lorentz boost.

\end{document}